\documentclass{ws-ijmpcs}

\newcommand\etal {{\it et al.}}

\newcommand\al{\alpha}

\newcommand\ga{\gamma}
\newcommand\de{\delta}

\newcommand\la{\lambda}

\newcommand\si{\sigma}

\newcommand\ps{\psi}

\newcommand\Ga{\Gamma}

\newcommand\Si{\Sigma}

\newcommand\mn{{\mu\nu}}

\newcommand\fr[2]{{{#1} \over {#2}}}
\newcommand\half{{\textstyle{1\over 2}}}

\newcommand\fracn[2]{{\textstyle{{#1}\over {#2}}}}

\newcommand\thpr{{these proceedings}}

\newcommand\pt[1]{\phantom{#1}}
\newcommand\ol[1]{\overline{#1}}

\newcommand\vb[2]{e_{#1}^{{\pt{#1}}#2}}
\newcommand\ivb[2]{e^{#1}_{{\pt{#1}}#2}}
\newcommand\uvb[2]{e^{#1#2}}

\newcommand\ab{\overline{a}{}}

\newcommand\cb{\overline{c}{}}

\newcommand\eb{\overline{e}{}}

\newcommand\mt{m^{\rm T}}
\newcommand\ms{m^{\rm S}}

\newcommand\afb{(\ab_{\rm{eff}})}

\newcommand\afbx[1]{(\ab^{#1}_{\rm{eff}})}
\newcommand\cbx[1]{(\cb^{#1})}

\newcommand\afbe{\afbx{e}}

\newcommand\cbw{\cbx{w}}
\newcommand\afbw{\afbx{w}}

\newcommand\lrpartial{\raise 1pt\hbox{$\stackrel\leftrightarrow\partial$}}

\newcommand\lrDmu{\stackrel{\leftrightarrow}{D_\mu}}

\newcommand\atext{$a_\mu$}
\newcommand\btext{$b_\mu$}
\newcommand\ctext{$c_{\mu\nu}$}
\newcommand\dtext{$d_{\mu\nu}$}
\newcommand\etext{$e_\mu$}
\newcommand\ftext{$f_\mu$}
\newcommand\gtext{$g_{\la\mu\nu}$}
\newcommand\Htext{$H_{\mu\nu}$}

\newcommand\G{G_N}

\newcommand{\beq}{\begin{equation}}
\newcommand{\eeq}{\end{equation}}
\newcommand{\bea}{\begin{eqnarray}}
\newcommand{\eea}{\end{eqnarray}}
\newcommand{\bit}{\begin{itemize}}
\newcommand{\eit}{\end{itemize}}

\begin{document}

\markboth{Jay D.\ Tasson}
{Gravity Effects on Antimatter in the Standard-Model Extension}

%
%

\title{GRAVITY EFFECTS ON ANTIMATTER
IN THE STANDARD-MODEL EXTENSION}

\author{JAY D.\ TASSON}

\address{Department of Physics and Astronomy,
Carleton College,
Northfield, Minnesota 55057 USA\\
(Present address: Physics Department, St.\ Olaf College,
Northfield, Minnesota 55057 USA)\\
tasson1@stolaf.edu}

\maketitle

\begin{history}

\end{history}

\begin{abstract}
The gravitational Standard-Model Extension (SME) 
is the general field-theory based
framework for the analysis of CPT and Lorentz violation.
In this work we summarize the implications of Lorentz and CPT violation
for antimatter gravity in the context of the SME.
Implications of various attempts to place indirect limits
on anomalous antimatter gravity are considered in the context
of SME-based models.
\end{abstract}

\ccode{PACS numbers:11.30.Cp \and 04.80.Cc \and 11.30.Er}

\section{Introduction}
\label{intro}

Antimatter physics
is an area in which many predictions of our current best theories,
the Standard Model of particle physics and General Relativity
remain unverified.
Thus experiments with antimatter
provide the opportunity
to place these theories
on a stronger experimental foundation.
One aspect of both of our existing theories
that can be tested with antimatter
is Lorentz symmetry,
along with the associated CPT symmetry.\cite{cpt}
Beyond improving the foundation
of our existing theories,
the search for Lorentz and CPT violation
offers the potential to detect Planck-scale physics.\cite{jtrev}
Standard lore holds that our current theories
are the low-energy limit of a more fundamental theory.
Lorentz violation has been shown to arise
in some candidates for the underlying theory
including string theory scenarios\cite{ksp,hashimoto}
and others,
thus providing a means of searching for Planck-scale physics
with current technology.\cite{ksp}

A comprehensive field-theoretic framework
for investigating Lorentz and CPT symmetry
as an expansion about known physics 
is provided by the SME.\cite{ck,akgrav,nonmin}
The SME is not a specific model,
but a comprehensive test framework
containing known physics and having the power to predict the outcome of relevant experiments
that is 
ideally suited for a broad search.
These predictions are then compared with experimental results.
Since no compelling evidence for Lorentz or CPT violation has been found
to date,
a broad and systematic search 
may offer a more efficient way of seeking such violations
than the consideration of many specific models.
With this philosophy,
a few models that illustrate aspects of the general framework
are useful;
however, aggressive model building
is avoided until new physics is found.
This proceedings contribution
reviews SME-based work in the context of gravitational experiments
as well as an SME-based model that illustrates several possibilities
in antimatter gravity.

\section{Basics}
\label{theory}

The action for the QED-extension limit of the gravitationally coupled SME,\cite{akgrav}
\beq
S = S_\ps + S_{\rm gravity} + S_A,
\label{action}
\eeq
provides the basic theory relevant for the discussion to follow.
From left to right the partial actions
are the gravitationally coupled fermion sector,
the pure-gravity sector,
and the photon sector.
Each term consists of known physics
along with all Lorentz-violating terms that can be constructed
from the associated fields.
Since they are not directly relevant for the discussion to follow,
the explicit forms of $S_{\rm gravity}$ and $S_A$ are omitted here,
though in general they are of considerable interest
and have been the subject of a large number of tests.\cite{tables}
Here
we specialize to the popular minimal-SME limit,
involving operators of mass dimension 3 and 4
where the fermion-sector action takes the form:
\beq
S_\ps = 
\int d^4 x (\half i e \ivb \mu a \ol \ps \Ga^a \lrDmu \ps 
- e \ol \ps M \ps),
\label{fermion}
\eeq
with
\bea
\Ga^a
&\equiv & 
\ga^a - c_{\mu\nu} \uvb \nu a \ivb \mu b \ga^b
- d_{\mu\nu} \uvb \nu a \ivb \mu b \ga_5 \ga^b
\nonumber\\
&&
- e_\mu \uvb \mu a 
- i f_\mu \uvb \mu a \ga_5 
- \half g_{\la\mu\nu} \uvb \nu a \ivb \la b \ivb \mu c \si^{bc}, \\
M
&\equiv &
m + a_\mu \ivb \mu a \ga^a 
+ b_\mu \ivb \mu a \ga_5 \ga^a 
+ \half H_{\mu\nu} \ivb \mu a \ivb \nu b \si^{ab}.
\label{mdef}
\eea
Here \atext, \btext, \ctext, \dtext, \etext, \ftext, \gtext, \Htext\ 
are coefficient fields for Lorentz violation
and gravitational couplings
occur via the vierbein $\vb \mu a$
and the covariant derivative.
The Minkowski-spacetime fermion-sector limit
can be recovered via $\vb \mu a \rightarrow \de^a_\mu$.

The content of the coefficient fields 
can be understood in two ways:
explicit Lorentz violation and spontaneous Lorentz violation.
Explicit Lorentz violation
involves the specification of the content of the coefficient fields 
as an external choice,
whereas spontaneous Lorentz violation
involves dynamical coefficient fields 
that receive vacuum expectation values via 
the spontaneous breaking of Lorentz symmetry.
Spontaneous Lorentz violation
is analogous to the spontaneous breaking of $SU(2) \times U(1)$
symmetry in the Standard Model;
however,
unlike electroweak-symmetry breaking,
the vacuum values that arise
are vector or tensor objects
known as coefficients for Lorentz violation
that can be thought of as establishing preferred directions in spacetime.
In nongravitational experiments analyzed in Minkowski spacetime
seeking effects associated with the vacuum values,
the distinction between explicit and spontaneous Lorentz-symmetry breaking
is not relevant,
and spacetime-independent coefficients for Lorentz violation are typically assumed.
This assumption could be regarded as a leading term 
in an expansion of a more general function.
Energy and momentum conservation is also preserved in this limit.

It has been shown that explicit Lorentz violation
is typically incompatible with the Riemann geometry 
of existing gravity theories.\cite{akgrav}
This suggests that consideration of Lorentz violation in a gravitational context
should either be done in the context of a more general geometry\cite{akfin}
or one should specialize to the case of spontaneous breaking.
Here we consider the latter case.
As in Minkowski-spacetime work,
we consider spacetime-independent vacuum values,
but geometric consistency requires consideration of
certain contributions to the fluctuations about the vacuum values as well.

\section{Gravitational Tests}
\label{gt}

In a gravitational context,
Lorentz-violating effects
can stem from the pure-gravity sector
or gravitational couplings in other sectors.
The framework for post-newtonian experimental searches in the pure-gravity sector
is developed in Ref.\ \refcite{lvpn},
and numerous searches have been performed and proposed.\cite{gexpt}
Reference \refcite{lvgap}
provides an analysis of gravitational couplings in the fermion sector
including a detailed analysis of the experimental and observational
implications of spin-independent coefficient fields
\atext, \ctext,\ and \etext.
The vacuum values associated with these coefficient fields are denoted $\afb_\mu$,
for the countershaded 
(observable only under special circumstances) combination\cite{akjt}
$\ab_\mu - m \eb_\mu$,
and $\cb_\mn$ for the vacuum value associated with \ctext. 
These vacuum values correspond to the coefficients for Lorentz violation
discussed in Minkowski spacetime.
The experimental implications of these fermion-sector coefficients,
including those relevant for antimatter,
are reviewed here.
Additional work on spin couplings has also been done.\cite{gspin}

Sensitivity to 
coefficients $\afb_\mu$ and $\cb_\mn$
can be achieved via a wide variety of gravitational experiments\cite{lvgap}
including gravimeter experiments,\cite{baumann}
tests of the universality of free fall,\cite{bke,uff}
redshift tests,\cite{red}
spin-precession tests,\cite{jtspin}
experiments with devices traditionally used for short-range gravity tests,\cite{db}
and
solar-system tests.
In laboratory tests,
the key point
is that the above coefficients generate tiny corrections
to the gravitational force
both along and perpendicular to the usual free-fall trajectory
near the surface of the Earth.
The coefficients also alter the effective inertial mass of a test body
in a direction-dependent way
resulting in a nontrivial relation between force and acceleration.\cite{berschinger}
These effects are time dependent varying
at the annual and sidereal frequencies,
and
may also be particle-species dependent.
These properties lead naturally
to a 4-category classification
of 
laboratory tests that use Earth as a source.
Measurements of the coefficients for Lorentz violation
can be made by monitoring the gravitational acceleration or force over time,
which constitute
free-fall gravimeter tests
or force-comparison gravimeter tests
respectively.
Similarly the relative acceleration of,
or relative force on,
a pair of test bodies may be monitored
resulting in free-fall or force-comparison Weak Equivalence Principle 
(WEP) tests
respectively.

While the above tests with ordinary, neutral matter
yield numerous sensitivities to Lorentz violation,
versions
of the tests highlighted above
performed with  antimatter, charged particles,
and second- and third-generation particles
can yield sensitivities to Lorentz and CPT violation 
that are otherwise impossible or difficult to achieve. 
Reference \refcite{lvgap} considers gravitational experiments with antihydrogen,\cite{amol2013,aegis,gbar,hbarai,muelcharge,gstate,weax,quint}
charged-particle interferometry,\cite{chargeai,muelcharge}
ballistic tests with charged particles,\cite{charge}
and signals in muonium free fall.\cite{muon}
Positronium may also offer an interesting possibility.\cite{positronium}
Here we consider antimatter tests further.
A recent direct measurement of the fall of antihydrogen by the ALPHA collaboration
has generated an initial direct limit on differences in the free-fall rate
of matter and antimatter.\cite{amol2013}
Improved measurements are in preparation or have been suggested,
including tests using
a Moir\'e accelerometer,\cite{aegis}
trapped antihydrogen,\cite{gbar}
antihydrogen interferometry,\cite{hbarai,muelcharge}
gravitational quantum states,\cite{gstate}
and tests in space.\cite{weax}
Such experiments could obtain special sensitivities
to the SME coefficients $\afbw_\mu$ and $\cbw_\mn$
since
the sign of $\afbw_\mu$ reverses under CPT,
while the sign of $\cbw_\mn$ does not.
Hence
antimatter experiments
could place cleaner constrains
on certain combinations of SME coefficients
than can be obtained with matter
and could in principle observe novel behaviors
stemming from Lorentz violation in the SME.

\section{Isotropic Parachute Model}

Beyond providing a framework 
for the analysis of antimatter gravity experiments,
the general field-theoretic approach of the SME
illuminates some aspects of attempts 
to place indirect limits on the possibility of unconventional 
antimatter gravity.\cite{mntg}
Consideration of toy-model limits
of the SME such as the isotropic `parachute' model (IPM),\cite{lvgap} 
can help facilitate the discussion.
The IPM is constructed
by restricting the classical nonrelativistic Lagrange density of the SME  
in the Sun-centered frame $S$,
to the limit in which the only nonzero coefficients are
$\afbw_T$ and isotropic $\cbw_{\Si\Xi}$.
For a test particle T moving in the gravitational field
of a source S
the effective classical Lagrangian 
in this limit
can be written in the suggestive form
\beq
L_{\rm IPM} = \half \mt_i v^2 + \fr{\G \mt_g \ms_g}{r},
\eeq
where $v$ is the velocity,
$r$ is distance from the source,
$\mt_i$ is the effective inertial mass of T,
and $\mt_g$ and $\ms_g$ 
are the effective gravitational masses 
of T and S, respectively.
The effective masses are defined in terms of 
the coefficients for Lorentz violation 
and the conventional Lorentz invariant body masses $m^{\rm B}$ as follows:
\bea
\nonumber
m^{\rm B}_i &=& 
m^{\rm B} + \sum_w \fracn53 (N^w+N^{\bar{w}}) m^w \cbw_{TT} \\
m^{\rm B}_g &=& 
m^{\rm B} + \sum_w \Big( (N^w+N^{\bar{w}}) m^w \cbw_{TT}
+ 2 \al (N^w-N^{\bar{w}}) \afbw_T \Big).
\label{friedmasses}
\eea
Here B is T or S,
$N^w$ and $N^{\bar{w}}$
are the number of particles and antiparticles of type $w$,
respectively,
and $m^w$ is the mass of a particle of type $w$.

The defining conditions of the IPM for electrons, protons, and neutrons,
is
\beq
\al \afbw_T = \fracn 13 m^w \cbw_{TT},
\eeq
where $w$ ranges over $e$, $p$, $n$.
The conditions
result in equal gravitational and inertial masses
for a matter body B,
$m^{\rm B}_i = m^{\rm B}_g$,
which implies that no Lorentz-violating effects 
appear in gravitational tests to third post-newtonian order
using ordinary matter.
In contrast,
for an antimatter test body T, 
$m^{\rm T}_i \neq m^{\rm T}_g$
within the IPM, 
which implies observable signals may arise in comparisons
of the gravitational responses 
of matter and antimatter
or of different types of antimatter.
The following paragraphs
consider the implications of some typical arguments against
anomalous antimatter gravity for the IPM
as well as some new indirect limits.

As a first classic argument,
we consider the question of whether energy remains conserved
when matter and antimatter have different gravitational responses.\cite{pm}
This argument is likely moot
in the present SME-based discussion
since conservation of energy and momentum 
play a starring role in developing the model.\cite{lvgap}
Still, consideration of the classic thought experiment 
in which a particle and an antiparticle
are lowered in a gravitational field,
converted to a photon pair,
raised to the original location,
and reconverted to the original particle-antiparticle pair
is interesting.
Here one normally assumes,
for example,
that the particle-antiparticle pair
gain a particular amount of energy from the gravitational field as they fall,
and
this energy is converted to a pair of photons with no additional change
in gravitational field energy.
The photons are assumed to couple differently to gravity
than the particle-antiparticle pair,
and hence they lose a different amount of energy on the way back to the original height
than that gained by the particle-antiparticle pair on the way down,
resulting in a violation of conservation of energy.
To explore these issues in the IPM,
we first note that in the analysis of Ref.\ \refcite{lvgap},
the photons are conventional,
partly via an available coordinate choice.
Then we note that the CPT-odd coefficient $\afbw_T$
shifts the effective gravitational coupling of the particles
and the antiparticles relative to the photons by equal and opposite amounts.
This implies no net difference for the particle-antiparticle combination
and the photons occurs as a result of $\afbw_T$.
The role of the $\cb_\mn$ coefficient
appears to challenge the assumption of no change in the gravitational
field energy
as the particle-antiparticle pair converts to photons.
If the two systems have different gravitational couplings
such that they exchange different amounts of energy with the gravitational field
during their trips,
the field energy will also change as the coupling changes during the reaction.
Hence differing gravitational couplings
are not in conflict with energy conservation
when field energy is considered.

Neutral-meson systems
which provide natural interferometers
mixing particle and antiparticle states
provide another classic indirect argument against anomalous antimatter gravity.\cite{mlg}
These systems have already been used to place tight constraints 
on certain differences among the $\afbw_\mu$ coefficients
for $w$ ranging over quark flavors
via flat spacetime considerations.\cite{mesons,akmesons}
These limits imply no dominant constrains
for baryons,
which involve three valence quarks,
or for leptons
in the context of the IPM.
Moreover,
the tests involve
valence $s$, $c$, or $b$ quarks,
which are largely irrelevant for protons and neutrons.
The essential point is that the flavor dependence of Lorentz and CPT violation in the SME
implies that the IPM evades constraints from meson systems. 

A final popular argument
against anomalous antimatter gravity
considered following the construction
of the IPM in Ref.\ \refcite{lvgap},
is based on the binding energy content of 
baryons, atoms, and bulk matter.\cite{lis2}
A version of the argument 
relevant for the discussion of antihydrogen
could begin by noting
that the quarks in hydrogen contain less than about 10\% of the mass,
with much of the remainder contained 
in the gluon and sea binding.
It might then be concluded that 
the gravitational response of the two cannot differ 
by more than about 10\%
based on their comparable binding forces.
Such arguments
typically assume implicitly 
that the gravitational response of a body is determined
by its mass and hence by binding energy.
In the IPM,
the coefficient $\afbw_T$,
leads to a correction to the gravitational force
that is independent of mass,
but can vary with flavor.
Hence the modifications to the gravitational responses
are determined primarily by the flavor content
of the valence particles.
A scenario in which
the anomalous gravitational effect is associated
purely with the positron could even be considered,
as would occur in the IPM when $\afbe_T$ is the only nonzero coefficient.
An investigation of radiative effects
involving $\afbw_T$, $\cbw_{TT}$,
and other SME coefficients for Lorentz violation\cite{ck,renorm}
could result in more definitive statements along the above lines,
perhaps with the IPM condition imposed after renormalization;
however,
the key points illustrated with the IPM are:
the anomalous gravitational response of a body
can be independent of mass,
can vary with flavor,
and can differ between particles and antiparticles.

An argument against anomalous antimatter gravity
not previously considered in the context of the IPM
is based on treating the cyclotron frequency
of a trapped particle/antiparticle as a clock,
which could receive an anomalous redshift
in certain models with differing gravitational responses for matter
and antimatter.\cite{holz}
The basic idea is to assume equivalent frequencies
for the clock and anticlock far from the source of the gravitational potential,
constrain the difference in the frequencies in the lab,
and extract a constraint based on the difference in the gravitational potential.
Contributions from the CPT even $c_\mn$ coefficient
in the IPM
have no effect since they are
the same for a particle and the corresponding antiparticle.
Though the $\ab_\mu$ coefficient takes the opposite value for particles and antiparticles,
it does not typically enter the redshift
as can be seen in the example of the Bohr levels of hydrogen\cite{lvgap}
as well as in other systems that have been considered in this context.\cite{red}
Hence the IPM is not likely impacted by this argument.

Though the IPM is a field-theoretic toy model
generating an anomalous gravitation response for antimatter
that appears to evade many of the typical indirect limits,
the model can be limited by a rather different type of
investigation with matter.
Certain experiments with sensitivity to higher powers of velocity,\cite{lvgap}
including the recent redshift analysis 
of matter systems,\cite{red}
considerations of bound kinetic energy,\cite{bke}
and double-boost suppression terms,
if analyzed,
in some flat-spacetime tests
have or could constrain the IPM
below the sensitivity goals of many upcoming antimatter gravity experiments.
The best constraints at present
are based on bound kinetic energy
and limit the anomalous gravitational response
of antimatter in the IPM to parts in $10^8$.\cite{bke}
However it is important to note
that these constraints are quite different from many of the usual arguments
against anomalous antimatter gravity such as those noted above.
They involve different types of physical arguments
and different experimental systems
highlighting the freedom that may exist
in constructing models that are insensitive to the usual constraints.
Note also that these constraints are of immediate relevance
only to the IPM, a special toy-model limit of the SME.
The possibility of constructing models similar to the IPM
based on the recently analyzed higher-order terms in the SME\cite{nonmin}
remains.

\section{Summary}

The SME provides a general field-theoretic framework
for seeking Lorentz and CPT violation,
a search that probes Planck-scale physics
with existing technology.
The comparison of matter and antimatter
provides a means of conducting such tests,
and a special limit of the SME provides a field-theoretic toy model
for investigating indirect limits on antimatter gravity.

\end{document}